\documentclass[hyperref,amsmath,amssymb,showpacs,floatfix,journal=ancac3,manuscript=article]{achemso}
\usepackage{graphicx}	
\usepackage{amsmath}	
\usepackage{amssymb}	
\usepackage{extarrows}
\usepackage{morefloats}
\mciteErrorOnUnknownfalse 
\usepackage[version=4]{mhchem} 
\setkeys{acs}{etalmode=truncate, maxauthors=200}
\setkeys{acs}{keywords = true}
\setkeys{acs}{usetitle=true}
\usepackage{epstopdf} 
\usepackage[perpage]{footmisc}

\usepackage[final]{changes}

\DeclareGraphicsExtensions{.eps, .ps, .pdf}
\graphicspath{{./}{/home/ioan/floppy/astro/mnras/}{/home/ioan/floppy/astro/mnras/data/}{/home/ioan/floppy/astro/mnras/data_c4n-/}}

\newcommand{\gl}{equation} 
\newcommand{\gls}{equations} 
\newcommand{\Gl}{Equation}
 
\newcommand{\figname}{Figure~} 

\newcommand{\secname}{Section~} 
\newcommand{\secsname}{Sections~}

\usepackage{mfirstuc}
\MFUnocap{of}
\MFUnocap{that}
\MFUnocap{by}
\MFUnocap{to}
\MFUnocap{on}
\MFUnocap{and}
\MFUnocap{in}
\MFUnocap{the}
\MFUnocap{for}
\MFUnocap{with}
\SectionNumbersOn 
\author{Ioan B\^aldea}
\email{ioan.baldea@pci.uni-heidelberg.de}
\affiliation{Theoretische Chemie, Universit\"at Heidelberg, Im Neuenheimer Feld 229, D-69120 Heidelberg, Germany}

\title[What Can We Learn from the Time Evolution of COVID-19 Epidemic in Slovenia?]  
{What Can We Learn from the Time Evolution of COVID-19 Epidemic in Slovenia?}

\begin{document}


\begin{abstract}
  A recent work (DOI 10.1101/2020.05.06.20093310) indicated that 
  temporarily splitting larger populations into smaller groups can efficiently mitigate the spread of SARS-CoV-2
  virus. The fact that, soon afterwards, on May 15, 2020, the two million people Slovenia was the first European
  country proclaiming the end of COVID-19 epidemic within national borders may be relevant from this perspective.
  Motivated by this evolution, in this paper we investigate the time dynamics of coronavirus cases in Slovenia
  with emphasis on how efficient various containment measures act to diminish the number of COVID-19 infections.
  Noteworthily, the present analysis does not rely on any speculative theoretical assumption; it is solely based on raw epidemiological data.
  Out of the results presented here, the most important one is perhaps
  the finding that, while imposing drastic curfews and travel restrictions
  reduce the infection rate $\kappa$ by a factor of four with respect to
  the unrestricted state, they only improve the $\kappa$-value by $\sim 15$\%
  as compared to the much bearable state of social and economical life wherein
  (justifiable) wearing face masks and social distancing rules are enforced/followed.
  Significantly for behavioral and social science, our analysis of the time dependence $\kappa = \kappa(t)$
  may reveal an interesting self-protection instinct of the population, which became manifest
  even before the official lockdown enforcement.
\end{abstract}

\section{Introduction}
\label{sec:intro}
In the unprecedented difficulty created by the COVID-19 pandemic outbreak \cite{WHO_COVID-19_Pandemic},
mathematical modeling developed by epidemiologists over many decades
\cite{Kermack:27,Kermack:32,Kermack:33,Bailey:75,Hethcote:94,Hethcote:00}
may make an important contribution in helping politics
to adopt adequate regulations to efficiently fight against the spread of SARS-CoV-2 virus
while mitigating negative economical and social consequences.
The latter aspect is of paramount importance \cite{Baldwin:20} also because, if not
adequately considered by governments currently challenged to deciding possibly
under dramatic circumstances and formidable tight schedule,
it can jeopardize the healthcare system itself. As an effort in this direction,
we drew recently attention \cite{Baldea:2020e}
to the general fact that the spread of the SARS-CoV-2 virus in smaller groups
can be substantially slowed down as compared to the case of larger populations.
In this vein, the time evolution of COVID-19 disease in the two million people Slovenia
certainly deserves special consideration, as on
15 May 2020, concluding that this country has the best epidemic situation in Europe,
Prime Minister Janez Jan\v{s}ka declared the end of the COVID-19 epidemic within Slovenian borders
\cite{EndCOVID19-Slovenia}. 
Subsequent developments (only four new cases between 15 and 24 May \cite{Covid19Slovenia_wikipedia}, cf.~Table~\ref{table:slo})
have fortunately given further support
to this declaration.
Attempting to understanding and learning from this sui generis circumstance
is the very aim of the present paper.

Thanks to long standing efforts extending over many decades, a rich arsenal of theoretical methods
of analyzing epidemics exists. Most of them trace back to the celebrated SIR model
\cite{Kermack:27,Kermack:32,Kermack:33,Bailey:75,Hethcote:94,Hethcote:00}
wherein the time evolution of the numbers of individuals belonging to various epidemiological classes
(susceptible (S), infected (I), recovered (R), etc) classes is described by
deterministic differential equations. Unfortunately, those approaches need many input
parameters \cite{IJIMAI-3841,Hermanowicz2020.03.31.20049486}
that can often be reliably estimated only after epidemics ended \cite{RKI2Parameters:2020},
which unavoidably compromises their ability of making predictions.
As an aggravating circumstance, one should also add
the difficulty not encountered in the vast majority of previous studies:
how do the input parameters needed in model simulations change in time
under so many restrictive measures 
(wearing face masks, social distancing, movement restrictions, isolation and quarantine policies, etc)
unknown in the pre-COVID-19 era?
Estimating model parameters from data fitting in a certain time interval to make predictions
can easily run into a difficulty like that described in the first paragraph of
\secname\ref{sec:t-dependent-kappa}.

As shown below, our approach obviates the aforementioned difficulty.
We will adopt a logistic growth model in a form which is different from
that often employed 
in the past \cite{Verhulst:1838,Verhulst:1847,Quetelet:1848,Ostwald:1883,Waggoner:00}
(see \secsname\ref{sec:t-dependent-kappa} and \ref{sec:conclusion} for technical details).
This model is considerably simpler than SIR flavors, and already turned out to be
an appealing framework in dealing with current COVID-19 pandemic issues
\cite{Hermanowicz2020.03.31.20049486,Baldea:2020e}.
Logistic functions (see \gl~(\ref{eq-n(t)-tau}) below) were utilized for studying various problems  
\cite{McKendrick:1912,Lloyd:67,Cramer:02,Vandermeer:10,Baldea:2017m,Baldea:2018e}.
Studies on population dynamics of epidemic populations
\cite{Fracker:36,Large:45,Plank:60,Plank:63,Plank:66,Zadoks:88,Vanderplank:76}
were also frequently based on the logistic function.

Nevertheless, as anticipated, there is an important difference between
the present approach (\secname\ref{sec:t-dependent-kappa})
and all the other approaches of which we are aware. The latter merely justify the logistic model
by the fact that recorded disease numbers followed a sigmoidal curve.
Shortcomings of this standpoint are delineated in the beginning of \secname\ref{sec:t-dependent-kappa}.
The strength of the approach presented in \secname\ref{sec:t-dependent-kappa} is 
that we do not use data fitting. Rather, we use raw epidemiological data to validate
the logistic growth and straightforwardly extract the time dependent infection rate,
which is the relevant model parameter for the specific case considered
and makes it possible to compare how efficient different restrictive measures act to mitigate
the COVID-19 pandemic, and even to get insight significant for behavioral and social science.
\section{Results and Discussion}
\label{sec:results}
\subsection{Standard Logistic Model}
\label{sec:logistic-model}
To briefly remind, standard logistical growth in time $t$ of an infected population $n = n(t)$
follows an ordinary differential equation 
\begin{equation}
\label{eq-ode}
\frac{d}{d t} n = \kappa n \left(1 - \frac{n}{N}\right)
\end{equation}
containing two constants (input model parameters):
the (intrinsic) infection rate
$\kappa$($>0$) and the so called carrying capacity $N$.
In a given environment, the latter has a fixed value
to which the population saturates asymptotically ($\lim_{t\to \infty} n(t) = N$).
This can be seen by straightforwardly integrating \gl~(\ref{eq-ode}) 
\begin{equation}
\label{eq-n(t)-tau}
n(t) = \frac{N}{1 + \left(\frac{N}{n_0} - 1\right)
  e^{-\kappa \left(t - t_0 \right)}
}
= \frac{N}{1 + e^{-\kappa \left(t - \tau\right)}}
\end{equation}
with the initial condition $n(t)\vert_{t=t_0} = n_0$, which is often recast 
by using the half-time
$\tau \equiv t_{0} + \frac{1}{\kappa}\ln \left(\frac{N}{n_0} - 1\right) $,
$n(t)_{t=\tau} = N/2$. Noteworthily for the discussion that follows, \gl~(\ref{eq-n(t)-tau})
assumes time-independent model parameters.
\begin{figure*}
  \centerline{  
      \includegraphics[width=0.45\textwidth]{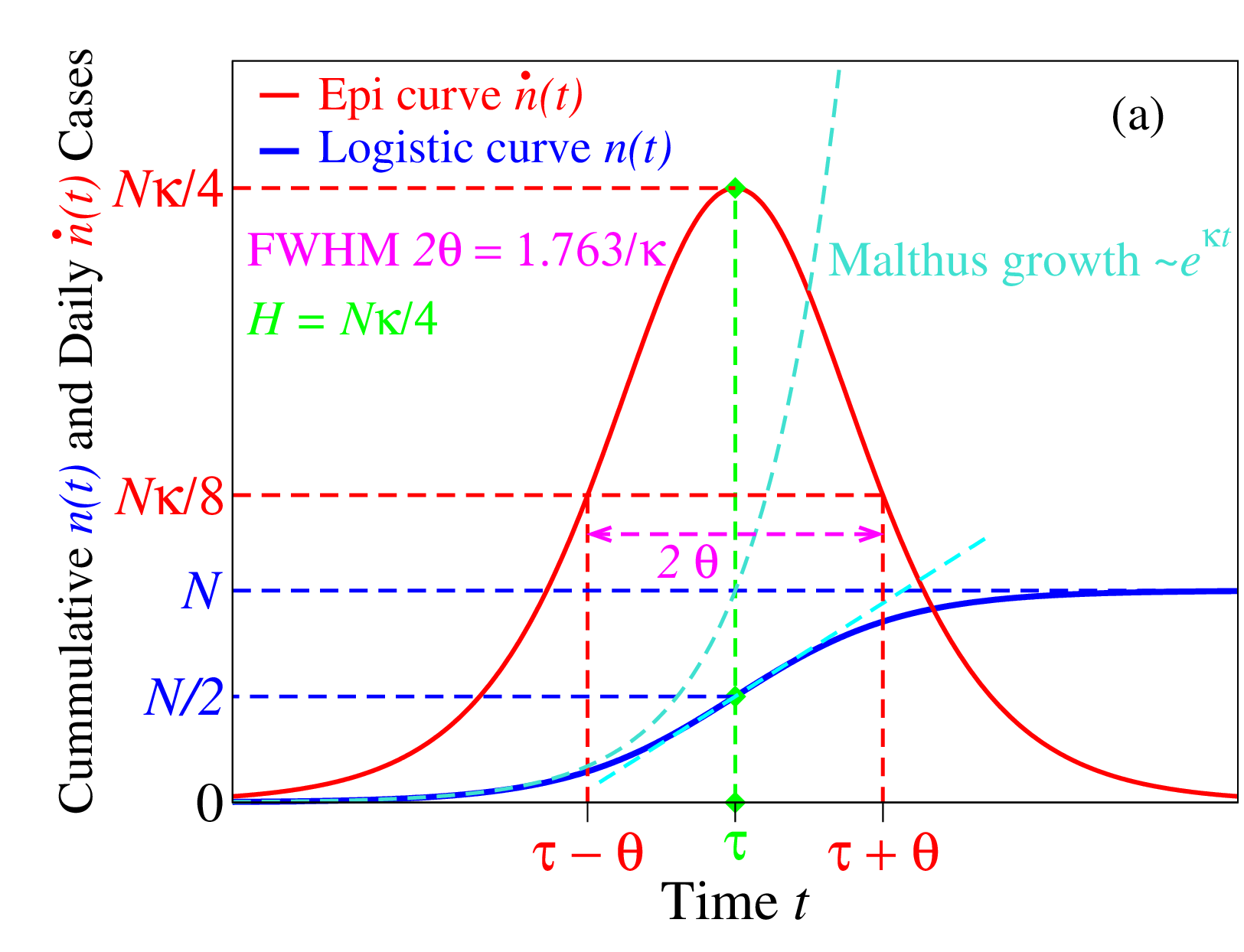}
         \includegraphics[width=0.45\textwidth]{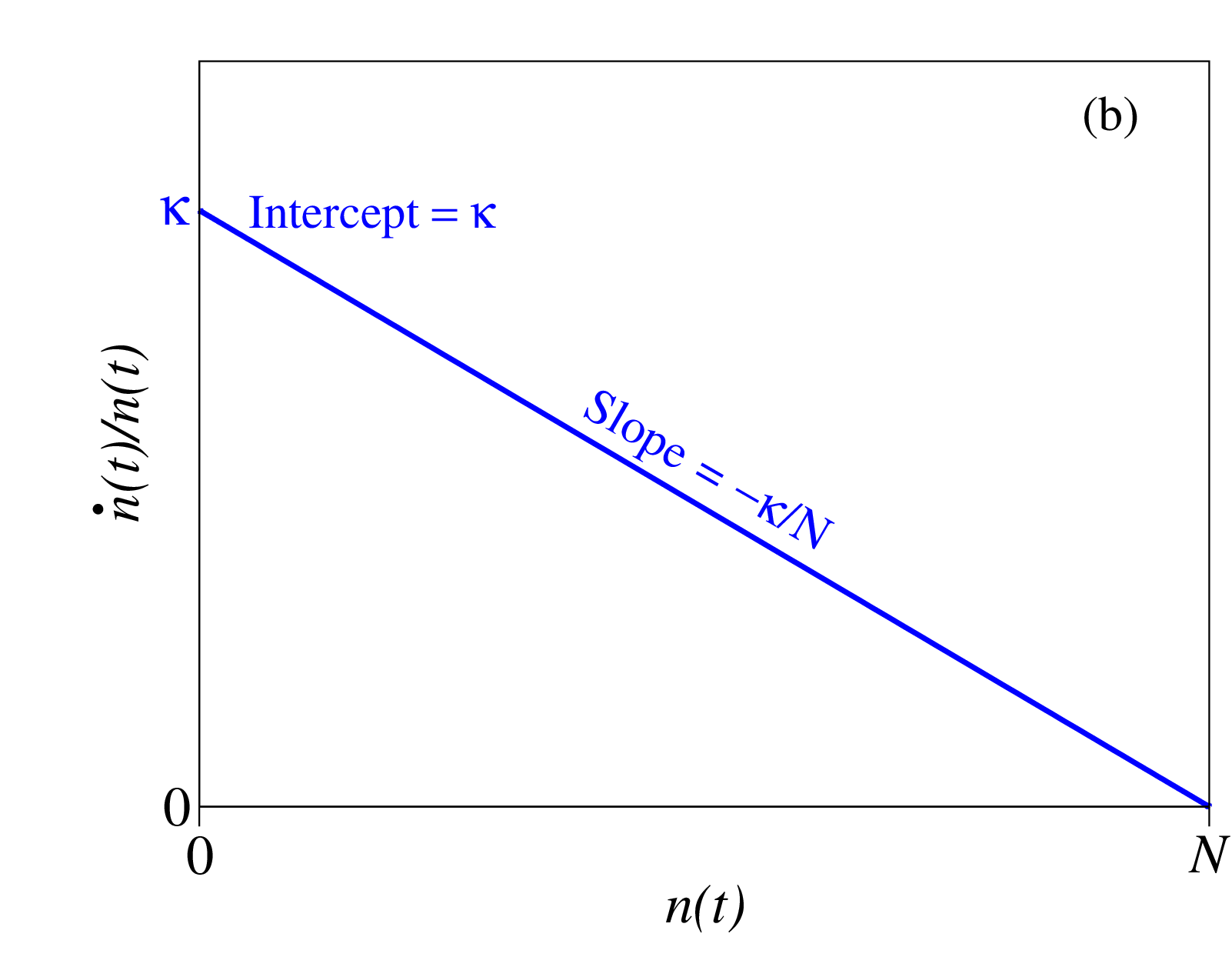}
  }      
  \caption{Panel a collects general properties of the standard logistic model. Panel b depicts a feature
  of the logistic growth whose importance is analyzed in \secname\ref{sec:t-dependent-kappa}.}
    \label{fig:generic}
\end{figure*}
In epidemiological language, $n(t)$ gives the cumulative number of cases
at time $t$. Plotted as a function of $t$, the derivative with respect to time
(throughout assumed a contiguous variable)
$\dot{n}(t) \equiv d n/ d t$, 
\begin{equation}
\label{eq-dn(t)/dt}
\dot{n}(t) = \frac{N\kappa}{4}\,\mbox{sech}^2\frac{\kappa (t - \tau)}{2} ,
\end{equation}
representing the ``daily'' number of new infections,
is referred to as the epi(demiological) curve.

\figname\ref{fig:generic}a summarizes in graphical form
the basic properties of logistic growth emerging the above equations.
The importance of the result presented in \figname\ref{fig:generic}b will become
apparent in \secname\ref{sec:t-dependent-kappa}.
\subsection{Brief COVID-19 Timeline in Slovenia}
\label{sec:slovenia}
Before proceeding with the data analysis
let us briefly summarize relevant public heath measures, social distancing and movement restrictions
imposed during the COVID-19 crisis 
in Slovenia \cite{Covid19Slovenia_wikipedia,Covid19Timeline}.

The first case of coronavirus was confirmed on March 4, 2020, imported via
a returnee traveling from Morocco via Italy
\cite{FirstCaseCovid19Slovenia}.
On 10 March, the government banned all incoming flights from
Italy, South Korea, Iran, and China; the land border with Italy was closed for all but freight transport;
indoor public with more than 100 persons were prohibited, sporting and other events with more than 500 participants
were allowed only without audience.

On 12 March,
the nationwide COVID-19 epidemic was proclaimed in Slovenia.
On 14 March the Crisis Management Staff of the Republic of Slovenia
established by the new government led by Prime Minister Janez Jan\v{s}a, confirmed on 13 March
amidst the coronavirus outbreak, suspended unnecessary services. On 15 March all restaurants and bars as well as
the Ljubljana Zoo were closed. On 16 March educational institutions, including kindergartens,
primary and secondary schools were closed down, and public (bus, rail, air) transport was stopped.
On 18 March public services (libraries, museums, cinemas, galleries were closed.
On 19 March public gatherings were limited; gatherings in higher
educational institutions and universities were prohibited; border checks (temperature, certificates of being healthy)
were introduced.
On 20 March de facto quarantine 
was established in Slovenia.

Significant easing occurred starting from 20 April.
Border crossing gradually reopened (6, 22 and 24 April).
Public transport resumed (27 April),
some pupils returned to schools (27 April), all bars and restaurants as well as small
hotels (up to 30 rooms) reopened for the first time since the shutdown was enforced in mid-March.
On 30 April, the general prohibition of movement was lifted.

Concluding that his country has the best epidemic situation in Europe,
Prime Minister Janez Jan\v{s}ka declared on 15 May the end of the COVID-19 epidemic within Slovenian borders
and allowed EU citizens free entrance.
\subsection{Logistic Model with Time Dependent Parameters}
\label{sec:t-dependent-kappa}
\begin{figure*}
  \centerline{  
    \includegraphics[width=0.3\textwidth]{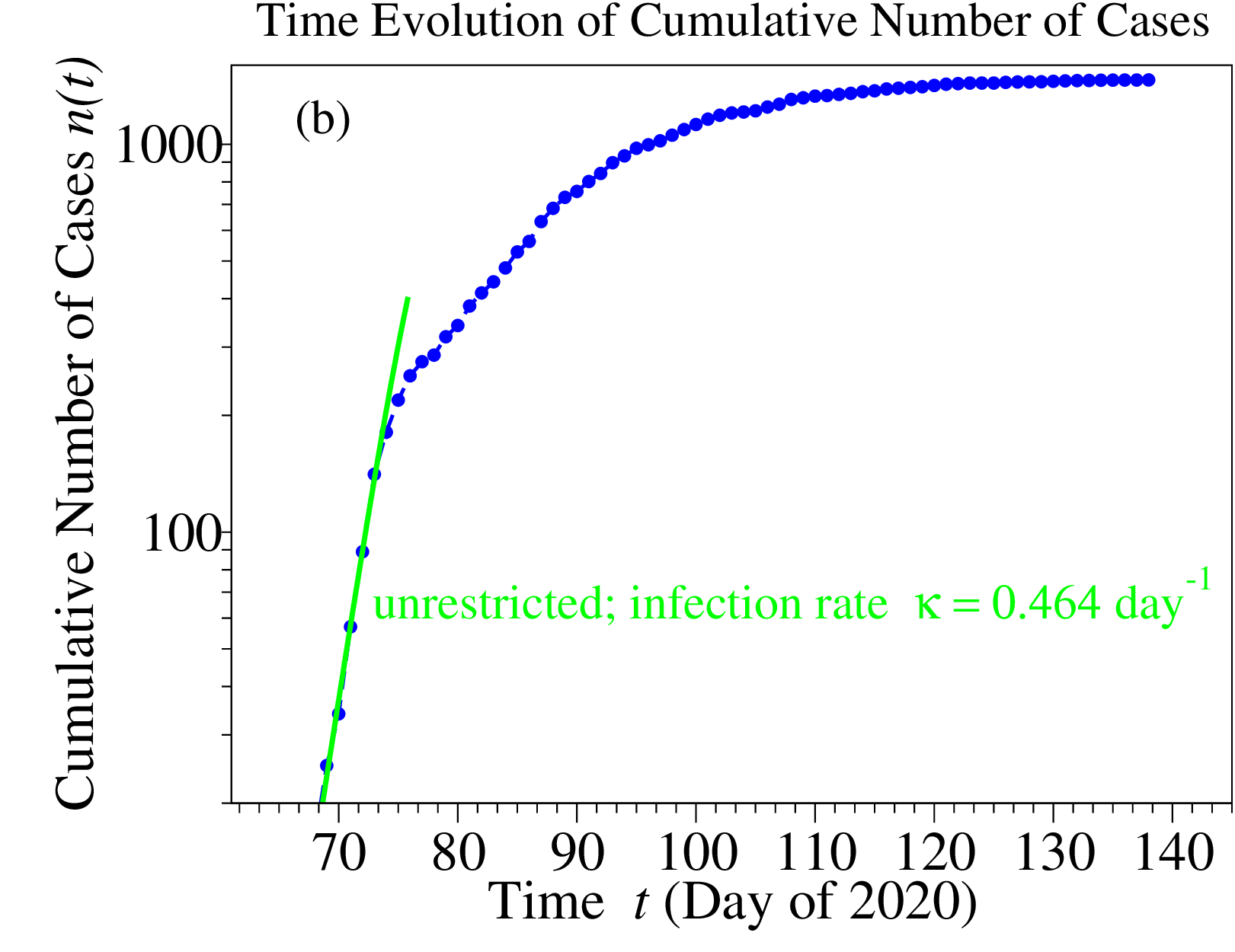}
    \includegraphics[width=0.3\textwidth]{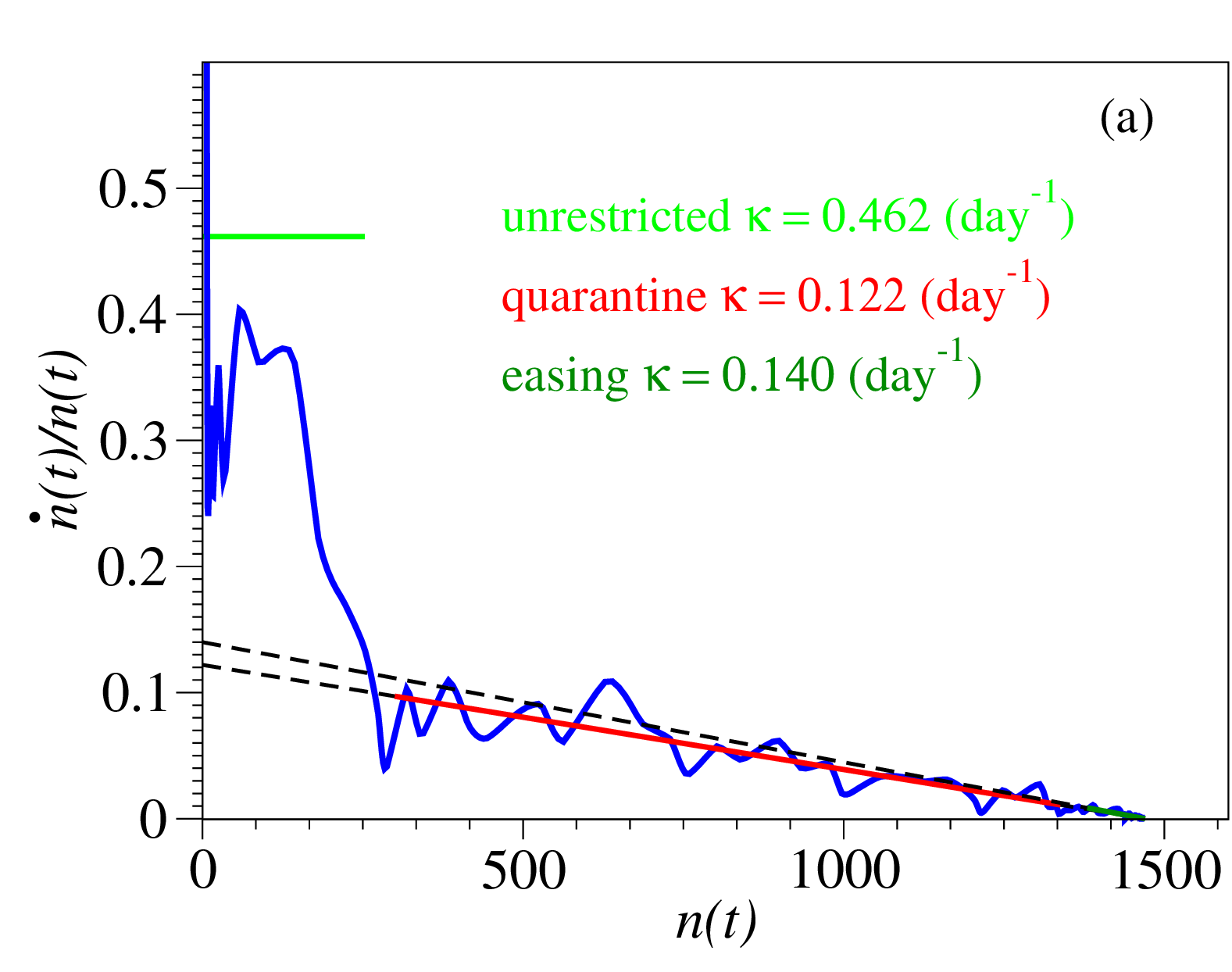}   
    \includegraphics[width=0.3\textwidth]{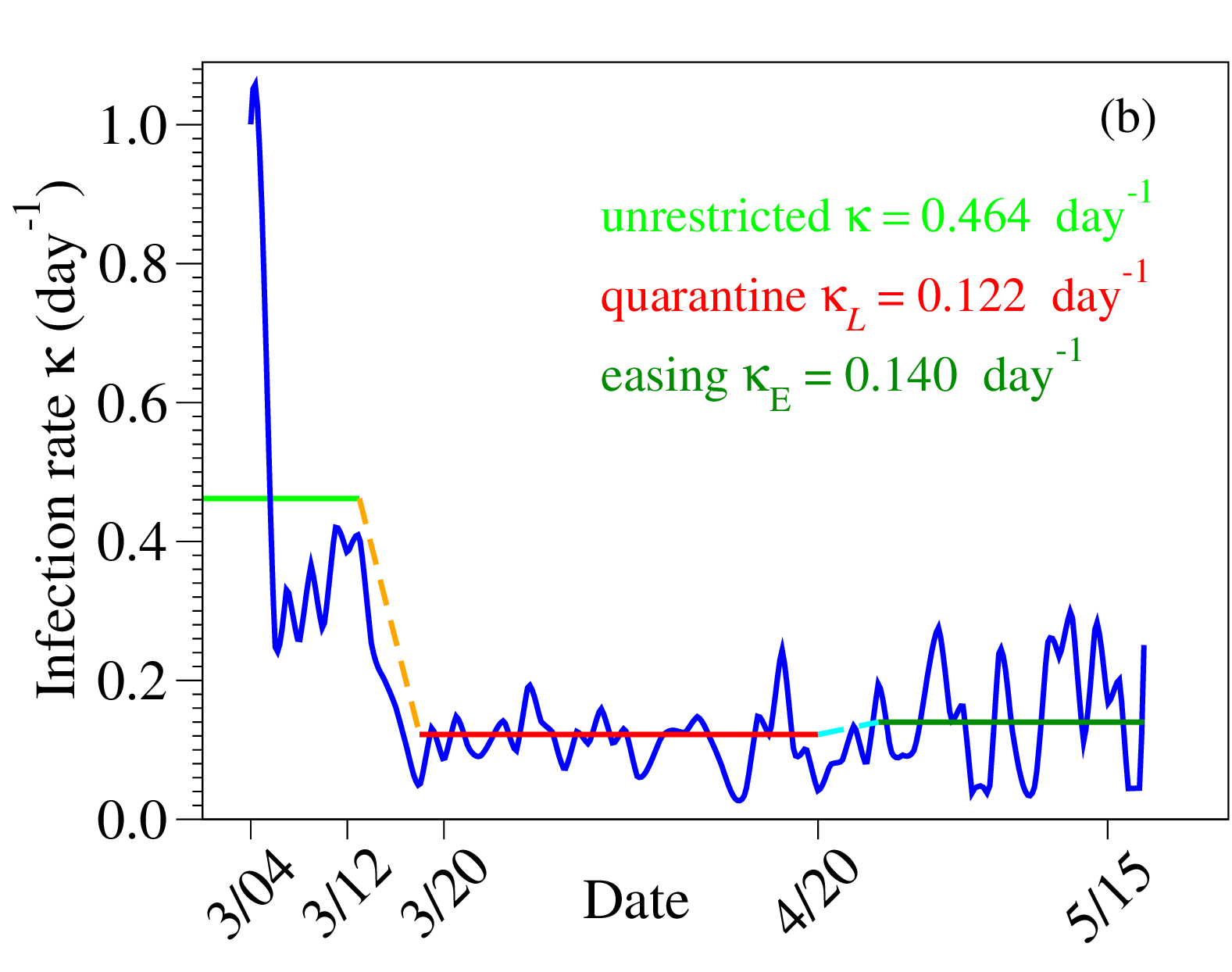}
  }  
  \caption{(a) In the short initial phase of the epidemic,
    the number of infected individuals growths exponentially in time,
    which allows to estimate the initial value of the infection rate ($\kappa = 0.464\,\mbox{day}^{-1}$.
    The curve of the logarithmic derivative $d \ln n(t)/d t = \dot{n}(t)/n(t)$ (blue line in panel b),
    which is merely obtained from epidemiological reports without any theoretical assumption,
    exhibits a linear trend that validates the description based on the logistic model. The carrying
    capacity $N$($=1470$) is obtained from the intersection with the x-axis.  
    The curve of the time-dependent infection rate $\kappa = \kappa(t)$ (blue line in panel c)
    deduced by means of \gl~(\ref{eq-kappa-ib}) suggests a piecewise linear approximation.
    Noteworthy, the infection rate $\kappa_{E} = 0.140\,\mbox{day}^{-1}$ after lockdown easing
    is only slightly larger than the value $\kappa_L = 0.122\,\mbox{day}^{-1}$ estimated for
    strict lockdown.}
    \label{fig:ib}       
\end{figure*}

The (blue) curve of \figname\ref{fig:ib}a depicting the evolution of total COVID-19
infections in Slovenia (underlying data are collected in Table~\ref{table:slo})
has an appealing similarity to the logistic S-shaped curve depicted in \figname\ref{fig:generic}.
One would be therefore tempting to follow 
numerous previous authors  
\cite{Fracker:36,Large:45,Plank:60,Plank:63,Plank:66,Zadoks:88,Vanderplank:76},
who claimed that the logistic model applies merely because of the (apparently) good data fitting.

Still, to claim that a description
based on a model like that of \gl~(\ref{eq-n(t)-tau}) is valid,
checking that the model parameters do not depend on the fitting range ($t_1 , t_2$)
is mandatory. For the specific case considered here, this means that
fitting numbers of infected individuals in time range $t_1 < t < t_2$ should yield,
within inherent statistical errors, values of $N$ and $\kappa$ independent of $t_1$ and $t_2$. 
And, like in other known cases \cite{Baldea:2015c,Baldea:2015g},
this is just the stumbling block for the logistic function approach delineated in
\secname\ref{sec:logistic-model}.
In particular, the infection rate $\kappa$ should not depend on how broad the
is the range ($t_1 , t_2$); however, we checked by straightforward numerical calculations
that it does.

Given the real epidemic timeline delineated in \secname\ref{sec:slovenia},
the infection rate \emph{must} indeed depend on time, $\kappa = \kappa(t)$.
If the contrary was true, all containment measures would be useless.
But when $\kappa$ depends on $t$, \gls~(\ref{eq-n(t)-tau}) and (\ref{eq-dn(t)/dt})
no longer apply; they were deduced by integrating out
\gl~(\ref{eq-ode}) assuming a time-independent $\kappa$.

Fortunately, rather than merely inquiring how good the
fitting curve based on \gl~(\ref{eq-n(t)-tau}) is,
we are able to directly check
(and demonstrate, see below) the validity of a
\emph{time-dependent} logistic model
merely based on the real epidemiological reports.
To this aim, we recast the differential \gl~(\ref{eq-ode}),
which is the basic definition of the logistic growth
(\emph{not} to be confused with the logistic function of \gl~(\ref{eq-n(t)-tau})),
as follows
\begin{equation}
  \label{eq-ib}
    \frac{\dot{n}}{n} = \kappa\left(1 - \frac{n}{N}\right)
\end{equation}

When put in this way, one can straightforwardly get insight in how to proceed.
One should plot the ratio of the daily new cases 
to the cumulative number of cases 
(numerator and denominator in \gl~(\ref{eq-ib}), respectively) 
as a function of the cumulative number of cases and inspect whether
the curve is linear or departs from linearity. Is the decrease linear
(like anticipated
in the ideal simulation presented in \figname\ref{fig:generic}b),
we have the demonstration that the logistic growth model applies.

The curve constructed 
as described above using the COVID-19 epidemic reports for Slovenia (Table~\ref{table:slo},
ref\citenum{Covid19Slovenia_wikipedia}) is depicted in \figname\ref{fig:ib}a.
As visible there, letting alone the 
strong fluctuations (possibly also due to the different methodology of reporting cases
\cite{Covid19Slovenia_wikipedia}) in the initial stage,
there is a transition from a high-$\kappa$ regime to a low-$\kappa$ regime. Definitely,
the COVID-19 restrictive measures worked. Interestingly (or, perhaps better, significantly),
the low-$\kappa$ regime appears to set in on 17-18 March ($n=275-286$), suggesting 
a population's prudent reaction even prior the official lockdown enforcement (20 March).
This is even more important given the fact that, in view of the finite incubation time ($\sim 5$ days),
reported cases pertain to infections that occurred earlier.

An overall linear trend is clearly visible in the low-$\kappa$ regime
of \figname\ref{fig:ib}a. \Gl~(\ref{eq-ib})
makes it then possible to estimate the carrying capacity $N$
from point where the (extrapolated) straight line
intersects the x-axis, while the intercept (or slope) can be used to
deduce the infection rate $\kappa$.

Noteworthily, the low-$\kappa$ regime comprises two periods: lockdown and
lockdown easing. To quantify the differences between these two periods,
we used \gl~(\ref{eq-ib}) to analyze the epidemiological data
from 18 March to 20 April ($78 \leq t \leq 111$)
and after 20 April ($t > 111$) separately. We could not find any significant difference
in $N$. With the value $N=1470$ (to be compared with the cumulative number of cases 1468 on 24 May)
in hands we have computed via \gl~(\ref{eq-ib})
the time dependence of infection rate
\begin{equation}
  \label{eq-kappa-ib}
    \kappa(t) = \frac{\dot{n}(t)/n}{1 - n(t)/N}
\end{equation}
A slight difference
could thus be obtained ($\kappa_L \simeq 0.122\,\mbox{day}^{-1}$ versus
$\kappa_E \simeq 0.140\,\mbox{day}^{-1}$), leading to the schematic
representation depicted in \figname\ref{fig:ib}b.

\begin{figure*}
  \centerline{  
     \includegraphics[width=0.45\textwidth]{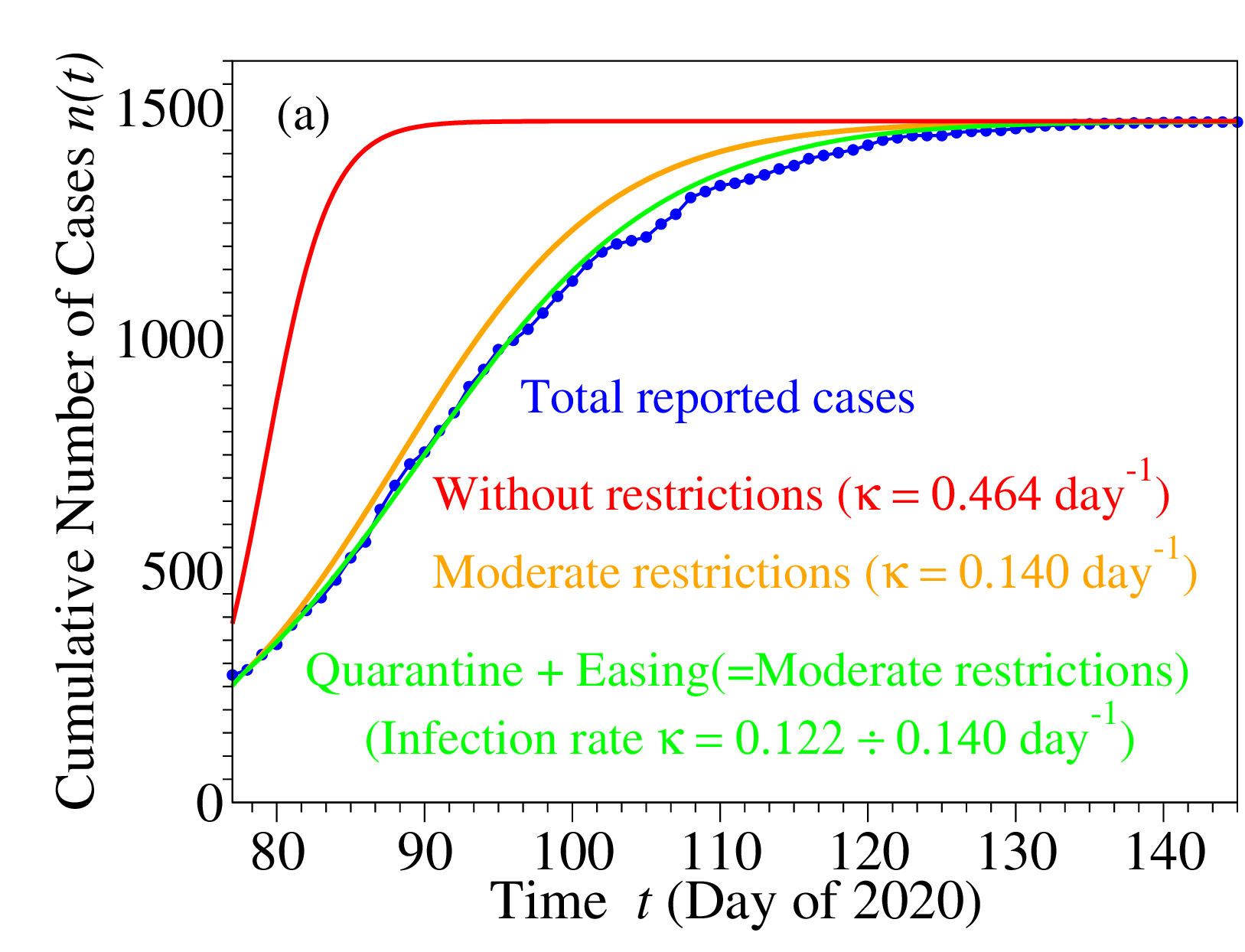}   
     \includegraphics[width=0.45\textwidth]{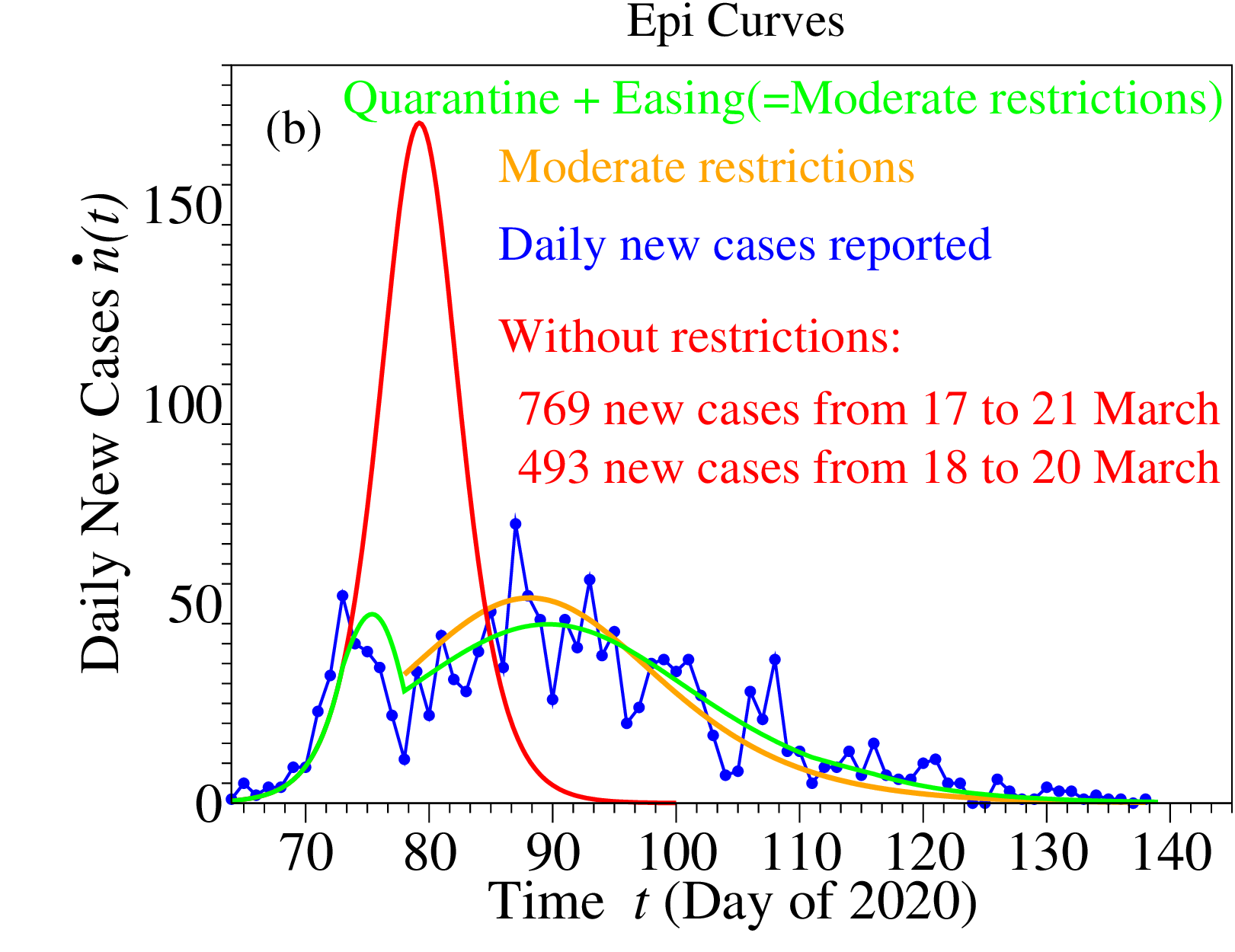}
  }  
  \caption{Curves for cumulative (panel a) and daily (panel b) number of cases based on epidemiological reports and simulated as
    described in the main text. Notice that the first peak around $t=73\,\mbox{day}^{-1}$
    does not represent a spurious effect of fluctuations. It is a real effect
    reproduced by numerical simulation (green curve), which nicely demonstrates how restrictive measures enforced
    stopped the explosive (unrestricted) evolution depicted in red. The orange curves depict results that could have been
    achieved if, instead of severe lockdown, the much milder restrictions effective on 20 April were imposed starting 20 March.}
    \label{fig:simulations}       
\end{figure*}

We used the piecewise linear approximation of \figname\ref{fig:ib}b
for the time dependent infection rate $\kappa \to \kappa(t)$ 
to numerically integrate the differential \gl~(\ref{eq-ode}).
Results of this numerical simulations for the total and daily number of cases are depicted
by the green curves in \figname\ref{fig:simulations} along with the blue curves representing
the epidemiological reports.

In addition, we also show there simulations of how the
Slovenian COVID-19 epidemic would have evolved if:

(a) no restriction measure had been enforced. The (red) curves indicate that the result could have been grim:
769 new infections in five days or 493 new cases in three days. Again, restriction measures were definitely necessary.

(b) instead of severe lockdown, much more bearable restrictive measures as effective on 20 April
would have been imposed on 20 March. The rather modest differences between the orange and green
curves suggest that, unless the healthcare system capacity is overwhelmed, 
(justified) face masks and social distancing rules can be comparably efficient in mitigating
the SARS-CoV-2 virus spread with draconian lockdown while obviating paralyzing economic and social life.
From this perspective, the ``German model'' appears to have an efficiency comparable
to that of the ``Italian model''.

\section{Conclusion}
\label{sec:conclusion}
We believe that especially the two results shown 
in \figname\ref{fig:ib}b ---
solely based on epidemiologically reported data
without any extra (possibly speculative) theoretical supposition ---
are of extreme practical importance, as they could help
setting adequate policies in the difficult period of the current COVID-19 pandemic:   
 
(i) Regarding the infection rate, it is worth emphasizing that,
while substantially smaller than without imposing
restrictions ($\kappa = 0.464\,\mbox{day}^{-1}$),
the value $\kappa_L = 0.122\,\mbox{day}^{-1}$ during shutdown
is only very slightly smaller than the value $\kappa_E = 0.140\,\mbox{day}^{-1}$.
after easing measures effective on 20 April. The important message
conveyed by this finding is that the efficiency of 
hardly bearable unselective quarantine and remain-isolated-at-home measures 
is very questionable. As one can intuitively expect, and what the present estimates
do quantify, is that what
really matters is not to keep everyone at home (``Italian approach'') but rather to
impede virus transmission (``German approach''),
e.g., by wearing masks, adequate hygiene, and social distancing.
Infection transmission does not strongly increase upon easing as long as face masks and social
distancing prevent SARS-CoV-2 virus spreading.
One should add at this point --- an important fact that appears to be currently
inadequately understand --- that, along with a less pleasant effect of a
short-term slight increase of the daily new cases, 
a moderate increase in the infection rate also has a positive impact.
It reduces epidemic duration; compare the \emph{right} tail 
the green and orange curves in \figname\ref{fig:simulations}b.

(ii) The fact that the carrying capacity $N$ does not change upon
lockdown easing is equally important. This is the maximum number of individuals that can be infected in
a given environment.
Rephrasing, the maximum number of infected individuals does not increase when the lockdown is released;
the total carrying capacity of a given environment does not change.

From a methodological perspective, one should emphasize the 
important technical strength of the approach proposed above, which made it possible
to arrive at the aforementioned conclusions.
It is only the \emph{differential} form, \gl~(\ref{eq-ode}), of logistic growth employed that obviates
the need for any additional theoretical assumption. The traditional approach
of validating the logistic model by blind data fitting using its \emph{integral} counterpart, 
\gl~(\ref{eq-n(t)-tau}), does not work for COVID-19 pandemic
applications because the model parameter $\kappa$ can and does depend on time.
This time dependence $\kappa = \kappa(t)$
is essential to properly assess and make recommendations on
the efficiency of the restriction measures
to be enforced against SARS-CoV-2 virus spread.

And just because, in its differential form utilized here,
the logistic model merely requires directly ``measurable'' epidemiological
quantities (daily reports $\dot{n}(t)$ and cumulative number of cases $n(t)$,
cf.~\gl~(\ref{eq-ode})) makes in the present unsusual situation
this model an alternative preferable to other more elaborate SIR-based flavors.
The latter models contain a series of quantities that cannot be directly accessed ``experimentally''.
Governments confronted to taking decisions under unprecedented time pressure
cannot await confirmation of
often speculative theoretical hypotheses needed in data processing.

Before ending, let us also note that monitoring the $\kappa(t)$-timeline
allowed us to get insight also relevant for behavioral and social science;
the self-protection instinct of the population became manifest
even before the official lockdown enforcement (cf.~\secname\ref{sec:t-dependent-kappa}).
%

\begin{mcitethebibliography}{37}
\providecommand*\natexlab[1]{#1}
\providecommand*\mciteSetBstSublistMode[1]{}
\providecommand*\mciteSetBstMaxWidthForm[2]{}
\providecommand*\mciteBstWouldAddEndPuncttrue
  {\def\EndOfBibitem{\unskip.}}
\providecommand*\mciteBstWouldAddEndPunctfalse
  {\let\EndOfBibitem\relax}
\providecommand*\mciteSetBstMidEndSepPunct[3]{}
\providecommand*\mciteSetBstSublistLabelBeginEnd[3]{}
\providecommand*\EndOfBibitem{}
\mciteSetBstSublistMode{f}
\mciteSetBstMaxWidthForm{subitem}{(\alph{mcitesubitemcount})}
\mciteSetBstSublistLabelBeginEnd
  {\mcitemaxwidthsubitemform\space}
  {\relax}
  {\relax}

\bibitem[Cucinotta and Vanelli(2020)Cucinotta, and
  Vanelli]{WHO_COVID-19_Pandemic}
Cucinotta,~D.; Vanelli,~M. WHO Declares COVID-19 a Pandemic. \emph{Acta Bio
  Medica Atenei Parmensis} \textbf{2020}, \emph{91}, 157--160\relax
\mciteBstWouldAddEndPuncttrue
\mciteSetBstMidEndSepPunct{\mcitedefaultmidpunct}
{\mcitedefaultendpunct}{\mcitedefaultseppunct}\relax
\EndOfBibitem
\bibitem[Kermack and McKendrick(1927)Kermack, and McKendrick]{Kermack:27}
Kermack,~W.~O.; McKendrick,~A.~G. Contributions to the mathematical theory of
  epidemics. I. \emph{Proc. Roy. Soc.} \textbf{1927}, \emph{115A}, 700--–721,
  reprinted in Bull. Math. Biol. 53, 33–55 (1991).
  https://doi.org/10.1007/BF02464423\relax
\mciteBstWouldAddEndPuncttrue
\mciteSetBstMidEndSepPunct{\mcitedefaultmidpunct}
{\mcitedefaultendpunct}{\mcitedefaultseppunct}\relax
\EndOfBibitem
\bibitem[Kermack and McKendrick(1932)Kermack, and McKendrick]{Kermack:32}
Kermack,~W.~O.; McKendrick,~A.~G. Contributions to the mathematical theory of
  epidemics—II. The problem of endemicity. \emph{Proc. Roy. Soc.}
  \textbf{1932}, \emph{138A}, 55--83, reprinted in Bull. Math. Biol. 53,
  57–87 (1991). https://doi.org/10.1007/BF02464424\relax
\mciteBstWouldAddEndPuncttrue
\mciteSetBstMidEndSepPunct{\mcitedefaultmidpunct}
{\mcitedefaultendpunct}{\mcitedefaultseppunct}\relax
\EndOfBibitem
\bibitem[Kermack and McKendrick(1933)Kermack, and McKendrick]{Kermack:33}
Kermack,~W.~O.; McKendrick,~A.~G. Contributions to the mathematical theory of
  epidemics. III. Further studies of the problem of endemicity. \emph{Proc.
  Roy. Soc.} \textbf{1933}, \emph{141A}, 94--122, reprinted in Bull. Math.
  Biol. 53, 89–118 (1991). https://doi.org/10.1007/BF02464425\relax
\mciteBstWouldAddEndPuncttrue
\mciteSetBstMidEndSepPunct{\mcitedefaultmidpunct}
{\mcitedefaultendpunct}{\mcitedefaultseppunct}\relax
\EndOfBibitem
\bibitem[Bailey(1975)]{Bailey:75}
Bailey,~N. T.~J. \emph{The Mathematical Theory of Infectious Diseases and Its
  Applications}; Charles Griffin \& Company Ltd, 5a Crendon Street, High
  Wycombe, Bucks HP13 6LE., 1975\relax
\mciteBstWouldAddEndPuncttrue
\mciteSetBstMidEndSepPunct{\mcitedefaultmidpunct}
{\mcitedefaultendpunct}{\mcitedefaultseppunct}\relax
\EndOfBibitem
\bibitem[Hethcote(1994)]{Hethcote:94}
Hethcote,~H.~W. A Thousand and One Epidemic Models. Frontiers in Mathematical
  Biology. Berlin, Heidelberg, 1994; pp 504--515\relax
\mciteBstWouldAddEndPuncttrue
\mciteSetBstMidEndSepPunct{\mcitedefaultmidpunct}
{\mcitedefaultendpunct}{\mcitedefaultseppunct}\relax
\EndOfBibitem
\bibitem[Hethcote(2000)]{Hethcote:00}
Hethcote,~H.~W. The Mathematics of Infectious Diseases. \emph{SIAM Review}
  \textbf{2000}, \emph{42}, 599--653\relax
\mciteBstWouldAddEndPuncttrue
\mciteSetBstMidEndSepPunct{\mcitedefaultmidpunct}
{\mcitedefaultendpunct}{\mcitedefaultseppunct}\relax
\EndOfBibitem
\bibitem[Baldwin and di~Mauro(2020)Baldwin, and di~Mauro]{Baldwin:20}
Baldwin,~R., di~Mauro,~B.~W., Eds. \emph{Mitigating the COVID Economic Crisis:
  Act Fast and Do Whatever It Takes}; A CEPR Press VoxEU.org eBook, 2020\relax
\mciteBstWouldAddEndPuncttrue
\mciteSetBstMidEndSepPunct{\mcitedefaultmidpunct}
{\mcitedefaultendpunct}{\mcitedefaultseppunct}\relax
\EndOfBibitem
\bibitem[B\^aldea(2020)]{Baldea:2020e}
B\^aldea,~I. Suppression of Groups Intermingling as Appealing Option For
  Flattening and Delaying the Epidemiologic Curve While Allowing Economic and
  Social Life at Bearable Level During COVID-19 Pandemic. \emph{medRxiv}
  \textbf{2020}, \relax
\mciteBstWouldAddEndPunctfalse
\mciteSetBstMidEndSepPunct{\mcitedefaultmidpunct}
{}{\mcitedefaultseppunct}\relax
\EndOfBibitem
\bibitem[End()]{EndCOVID19-Slovenia}
https://www.independent.co.uk/news/world/europe/oronavirus-slovenia-end-epidemic-lockdown-lifted-a9516841.html\relax
\mciteBstWouldAddEndPuncttrue
\mciteSetBstMidEndSepPunct{\mcitedefaultmidpunct}
{\mcitedefaultendpunct}{\mcitedefaultseppunct}\relax
\EndOfBibitem
\bibitem[Wikipedia()]{Covid19Slovenia_wikipedia}
Wikipedia, \url{https://en.wikipedia.org/wiki/COVID-19_pandemic_in_Slovenia},
  https://en.wikipedia.org/wiki/COVID-19\_pandemic\_in\_Slovenia\relax
\mciteBstWouldAddEndPuncttrue
\mciteSetBstMidEndSepPunct{\mcitedefaultmidpunct}
{\mcitedefaultendpunct}{\mcitedefaultseppunct}\relax
\EndOfBibitem
\bibitem[Fong \latin{et~al.}(2020)Fong, Li, Dey, Crespo, and
  Herrera-Viedma]{IJIMAI-3841}
Fong,~S.~J.; Li,~G.; Dey,~N.; Crespo,~R.~G.; Herrera-Viedma,~E. Finding an
  Accurate Early Forecasting Model from Small Dataset: A Case of 2019-nCoV
  Novel Coronavirus Outbreak. \emph{International Journal of Interactive
  Multimedia and Artificial Intelligence} \textbf{2020}, \emph{6},
  132--140\relax
\mciteBstWouldAddEndPuncttrue
\mciteSetBstMidEndSepPunct{\mcitedefaultmidpunct}
{\mcitedefaultendpunct}{\mcitedefaultseppunct}\relax
\EndOfBibitem
\bibitem[Hermanowicz(2020)]{Hermanowicz2020.03.31.20049486}
Hermanowicz,~S.~W. Simple model for Covid-19 epidemics - back-casting in China
  and forecasting in the US. \emph{medRxiv} \textbf{2020}, DOI
  10.1101/2020.03.31.20049486\relax
\mciteBstWouldAddEndPuncttrue
\mciteSetBstMidEndSepPunct{\mcitedefaultmidpunct}
{\mcitedefaultendpunct}{\mcitedefaultseppunct}\relax
\EndOfBibitem
\bibitem[Institut(2020)]{RKI2Parameters:2020}
Institut,~R.~K. Modellierung von Beispielszenarien der SARS-CoV-2-Epidemie 2020
  in Deutschland. 2020;
  \url{https://www.rki.de/DE/Content/InfAZ/N/Neuartiges_Coronavirus/Modellierung_Deutschland.pdf?__blob=publicationFile},
  https://www.rki.de/DE/Content/InfAZ/N/Neuartiges\_Coronavirus/Modellierung\_Deutschland.pdf?\_\_blob=publicationFile\relax
\mciteBstWouldAddEndPuncttrue
\mciteSetBstMidEndSepPunct{\mcitedefaultmidpunct}
{\mcitedefaultendpunct}{\mcitedefaultseppunct}\relax
\EndOfBibitem
\bibitem[Verhulst(1838)]{Verhulst:1838}
Verhulst,~P.-F. Notice sur la loi que la population poursuit dans son
  accroissement. \emph{Correspondance Math\'ematique et Physique}
  \textbf{1838}, \emph{10}, 113--121\relax
\mciteBstWouldAddEndPuncttrue
\mciteSetBstMidEndSepPunct{\mcitedefaultmidpunct}
{\mcitedefaultendpunct}{\mcitedefaultseppunct}\relax
\EndOfBibitem
\bibitem[Verhulst(1845)]{Verhulst:1847}
Verhulst,~P.-F. Recherches math\'emathiques sur la loi d'accroissement de la
  population. \emph{Nouveaux M\'emoires de l'Academie Royale des Sciences et
  Belles-Lettres de Bruxelles} \textbf{1845}, \emph{18}, 8\relax
\mciteBstWouldAddEndPuncttrue
\mciteSetBstMidEndSepPunct{\mcitedefaultmidpunct}
{\mcitedefaultendpunct}{\mcitedefaultseppunct}\relax
\EndOfBibitem
\bibitem[Quetelet(1848)]{Quetelet:1848}
Quetelet,~L. A.~J. \emph{Du Syst{\`e}me Social et des Lois qui le
  R{\'e}gissent}; Guillaumin, 1848\relax
\mciteBstWouldAddEndPuncttrue
\mciteSetBstMidEndSepPunct{\mcitedefaultmidpunct}
{\mcitedefaultendpunct}{\mcitedefaultseppunct}\relax
\EndOfBibitem
\bibitem[Ostwald(1883)]{Ostwald:1883}
Ostwald,~W. Studien zur chemischen Dynamik; Erste Abhandlung: Die Einwirkung
  der Säuren auf Acetamid. \emph{Journal f\"ur Praktische Chemie}
  \textbf{1883}, \emph{27}, 1--39\relax
\mciteBstWouldAddEndPuncttrue
\mciteSetBstMidEndSepPunct{\mcitedefaultmidpunct}
{\mcitedefaultendpunct}{\mcitedefaultseppunct}\relax
\EndOfBibitem
\bibitem[Waggoner and Aylor(2000)Waggoner, and Aylor]{Waggoner:00}
Waggoner,~P.~E.; Aylor,~D.~E. Epidemiology: A Science of Patterns. \emph{Annual
  Review of Phytopathology} \textbf{2000}, \emph{38}, 71--94, PMID:
  11701837\relax
\mciteBstWouldAddEndPuncttrue
\mciteSetBstMidEndSepPunct{\mcitedefaultmidpunct}
{\mcitedefaultendpunct}{\mcitedefaultseppunct}\relax
\EndOfBibitem
\bibitem[McKendrick and Pai(1912)McKendrick, and Pai]{McKendrick:1912}
McKendrick,~A.~G.; Pai,~M.~K. XLV. The Rate of Multiplication of
  Micro-organisms: A Mathematical Study. \emph{Proceedings of the Royal Society
  of Edinburgh} \textbf{1912}, \emph{31}, 649--653\relax
\mciteBstWouldAddEndPuncttrue
\mciteSetBstMidEndSepPunct{\mcitedefaultmidpunct}
{\mcitedefaultendpunct}{\mcitedefaultseppunct}\relax
\EndOfBibitem
\bibitem[Lloyd(1967)]{Lloyd:67}
Lloyd,~P. American, German and British antecedents to Pearl and Reed's logistic
  curve. \emph{Population Studies} \textbf{1967}, \emph{21}, 99--108\relax
\mciteBstWouldAddEndPuncttrue
\mciteSetBstMidEndSepPunct{\mcitedefaultmidpunct}
{\mcitedefaultendpunct}{\mcitedefaultseppunct}\relax
\EndOfBibitem
\bibitem[Cramer(2004)]{Cramer:02}
Cramer,~J. The early origins of the logit model. \emph{Studies in History and
  Philosophy of Science Part C: Studies in History and Philosophy of Biological
  and Biomedical Sciences} \textbf{2004}, \emph{35}, 613 -- 626\relax
\mciteBstWouldAddEndPuncttrue
\mciteSetBstMidEndSepPunct{\mcitedefaultmidpunct}
{\mcitedefaultendpunct}{\mcitedefaultseppunct}\relax
\EndOfBibitem
\bibitem[Vandermeer(2010)]{Vandermeer:10}
Vandermeer,~J. How Populations Grow: The Exponential and Logistic Equations.
  \emph{Nature Education Knowledge} \textbf{2010}, \emph{3}, 15\relax
\mciteBstWouldAddEndPuncttrue
\mciteSetBstMidEndSepPunct{\mcitedefaultmidpunct}
{\mcitedefaultendpunct}{\mcitedefaultseppunct}\relax
\EndOfBibitem
\bibitem[B\^aldea(2017)]{Baldea:2017m}
B\^aldea,~I. Floppy Molecules as Candidates for Achieving Optoelectronic
  Molecular Devices without Skeletal Rearrangement or Bond Breaking.
  \emph{Phys. Chem. Chem. Phys.} \textbf{2017}, \emph{19}, 30842 -- 30851\relax
\mciteBstWouldAddEndPuncttrue
\mciteSetBstMidEndSepPunct{\mcitedefaultmidpunct}
{\mcitedefaultendpunct}{\mcitedefaultseppunct}\relax
\EndOfBibitem
\bibitem[B\^aldea(2018)]{Baldea:2018e}
B\^aldea,~I. A sui generis electrode-driven spatial confinement effect
  responsible for strong twisting enhancement of floppy molecules in closely
  packed self-assembled monolayers. \emph{Phys. Chem. Chem. Phys.}
  \textbf{2018}, \emph{20}, 23492--23499\relax
\mciteBstWouldAddEndPuncttrue
\mciteSetBstMidEndSepPunct{\mcitedefaultmidpunct}
{\mcitedefaultendpunct}{\mcitedefaultseppunct}\relax
\EndOfBibitem
\bibitem[Fracker({1936})]{Fracker:36}
Fracker,~S. {Progressive intensification of uncontrolled plant-disease
  outbreaks}. \emph{{Journal of Economic Entomology}} \textbf{{1936}},
  \emph{{29}}, {923--940}\relax
\mciteBstWouldAddEndPuncttrue
\mciteSetBstMidEndSepPunct{\mcitedefaultmidpunct}
{\mcitedefaultendpunct}{\mcitedefaultseppunct}\relax
\EndOfBibitem
\bibitem[Large(1945)]{Large:45}
Large,~E.~C. Field trials of copper fungicides for the control of potato
  blight. I. Foliage protection and yield. \emph{Ann. Appl. Biol.}
  \textbf{1945}, \emph{32}, 319--329\relax
\mciteBstWouldAddEndPuncttrue
\mciteSetBstMidEndSepPunct{\mcitedefaultmidpunct}
{\mcitedefaultendpunct}{\mcitedefaultseppunct}\relax
\EndOfBibitem
\bibitem[van~der Plank(1960)]{Plank:60}
van~der Plank,~J.~E. In \emph{Plant Pathology}; Horsfall,~J.~G., Dimond,~A.~E.,
  Eds.; Academic Press, 1960; pp 229 -- 289\relax
\mciteBstWouldAddEndPuncttrue
\mciteSetBstMidEndSepPunct{\mcitedefaultmidpunct}
{\mcitedefaultendpunct}{\mcitedefaultseppunct}\relax
\EndOfBibitem
\bibitem[van~der Plank(1963)]{Plank:63}
van~der Plank,~J.~E. \emph{Plant Diseases: Epidemics and Control}; Academic,
  New York, 1963\relax
\mciteBstWouldAddEndPuncttrue
\mciteSetBstMidEndSepPunct{\mcitedefaultmidpunct}
{\mcitedefaultendpunct}{\mcitedefaultseppunct}\relax
\EndOfBibitem
\bibitem[Van Der~Plank(1966)]{Plank:66}
Van Der~Plank,~J.~E. Horizontal (polygenic) and vertical (oligogenic)
  resistance against blight. \emph{American Potato Journal} \textbf{1966},
  \emph{43}, 43 -- 52\relax
\mciteBstWouldAddEndPuncttrue
\mciteSetBstMidEndSepPunct{\mcitedefaultmidpunct}
{\mcitedefaultendpunct}{\mcitedefaultseppunct}\relax
\EndOfBibitem
\bibitem[Zadoks and Schein(1988)Zadoks, and Schein]{Zadoks:88}
Zadoks,~J.~C.; Schein,~R.~D. James Edward Vanderplank: Maverick and Innovator.
  \emph{Annual Review of Phytopathology} \textbf{1988}, \emph{26}, 31--37\relax
\mciteBstWouldAddEndPuncttrue
\mciteSetBstMidEndSepPunct{\mcitedefaultmidpunct}
{\mcitedefaultendpunct}{\mcitedefaultseppunct}\relax
\EndOfBibitem
\bibitem[Vanderplank(1976)]{Vanderplank:76}
Vanderplank,~J.~E. Four Essays. \emph{Ann. Rev. Phytopathol.} \textbf{1976},
  \emph{14}, 1--11, PMID: 22594585\relax
\mciteBstWouldAddEndPuncttrue
\mciteSetBstMidEndSepPunct{\mcitedefaultmidpunct}
{\mcitedefaultendpunct}{\mcitedefaultseppunct}\relax
\EndOfBibitem
\bibitem[Cov(2020)]{Covid19Timeline}
COVID-19: Government Measures Timeline. 2020;
  \url{https://gullivern.org/demo/metametis/covid-19/timeline_covid19_acaps.html},
  https://gullivern.org/demo/metametis/covid-19/timeline\_covid19\_acaps.html\relax
\mciteBstWouldAddEndPuncttrue
\mciteSetBstMidEndSepPunct{\mcitedefaultmidpunct}
{\mcitedefaultendpunct}{\mcitedefaultseppunct}\relax
\EndOfBibitem
\bibitem[Fir(2020)]{FirstCaseCovid19Slovenia}
2020; Reuters (4 March 2020). ``Slovenia Confirms First Case of Coronavirus:
  Health Minister''. The New York Times. ISSN 0362-4331. Retrieved 6 March 2020
  – via NYTimes.com.\relax
\mciteBstWouldAddEndPunctfalse
\mciteSetBstMidEndSepPunct{\mcitedefaultmidpunct}
{}{\mcitedefaultseppunct}\relax
\EndOfBibitem
\bibitem[B\^aldea(2015)]{Baldea:2015c}
B\^aldea,~I. Important Issues Facing Model-Based Approaches to Tunneling
  Transport in Molecular Junctions. \emph{Phys. Chem. Chem. Phys.}
  \textbf{2015}, \emph{17}, 20217--20230\relax
\mciteBstWouldAddEndPuncttrue
\mciteSetBstMidEndSepPunct{\mcitedefaultmidpunct}
{\mcitedefaultendpunct}{\mcitedefaultseppunct}\relax
\EndOfBibitem
\bibitem[B\^aldea(2015)]{Baldea:2015g}
B\^aldea,~I. Counterintuitive issues in the charge transport through molecular
  junctions. \emph{Phys. Chem. Chem. Phys.} \textbf{2015}, \emph{17},
  31260--31269\relax
\mciteBstWouldAddEndPuncttrue
\mciteSetBstMidEndSepPunct{\mcitedefaultmidpunct}
{\mcitedefaultendpunct}{\mcitedefaultseppunct}\relax
\EndOfBibitem
\end{mcitethebibliography}
\providecommand{\latin}[1]{#1}
\makeatletter
\providecommand{\doi}
  {\begingroup\let\do\@makeother\dospecials
  \catcode`\{=1 \catcode`\}=2 \doi@aux}
\providecommand{\doi@aux}[1]{\endgroup\texttt{#1}}
\makeatother
\providecommand*\mcitethebibliography{\thebibliography}
\csname @ifundefined\endcsname{endmcitethebibliography}
  {\let\endmcitethebibliography\endthebibliography}{}

\newpage
\begin{table} 
  \tiny 
  \begin{center}
    \begin{tabular*}{0.7\textwidth}{@{\extracolsep{\fill}}rrrr}
      \hline
Day of 2020 &    Date    &     Total Confirmed     &     New Confirmed per Day     \\
      \hline
   64   &   03-04    &           1             &       1                       \\
   65   &   03-05    &           6             &       5                       \\ 
   66   &   03-06    &           8             &       2                       \\
   67   &   03-07    &          12             &       4                       \\
   68   &   03-08    &          16             &       4                       \\
   69   &   03-09    &          25             &       9                       \\
   70   &   03-10    &          34             &       9                       \\
   71   &   03-11    &          57             &       23                      \\
   72   &   03-12    &          89             &       32                      \\
   73   &   03-13    &         141             &       52                      \\
   74   &   03-14    &         181             &       40                      \\
   75   &   03-15    &         219             &       38                      \\   
   76   &   03-16    &         253             &       34                      \\   
   77   &   03-17    &         275             &       22                      \\   
   78   &   03-18    &         286             &       11                      \\   
   79   &   03-19    &         319             &       33                      \\   
   80   &   03-20    &         341             &       22                      \\   
   81   &   03-21    &         383             &       42                      \\   
   82   &   03-22    &         414             &       31                      \\   
   83   &   03-23    &         442             &       28                      \\   
   84   &   03-24    &         480             &       38                      \\   
   85   &   03-25    &         528             &       48                      \\   
   86   &   03-26    &         562             &       34                      \\   
   87   &   03-27    &         632             &       70                      \\   
   88   &   03-28    &         684             &       52                      \\   
   89   &   03-29    &         730             &       46                      \\   
   90   &   03-30    &         756             &       26                      \\   
   91   &   03-31    &         802             &       46                      \\   
   92   &   04-01    &         841             &       39                      \\   
   93   &   04-02    &         897             &       56                      \\   
   94   &   04-03    &         934             &       37                      \\   
   95   &   04-04    &         977             &       43                      \\   
   96   &   04-05    &         997             &       20                      \\   
   97   &   04-06    &        1021             &       24                      \\   
   98   &   04-07    &        1056             &       35                      \\   
   99   &   04-08    &        1092             &       36                      \\   
  100   &   04-09    &        1125             &       33                      \\   
  101   &   04-10    &        1161             &       36                      \\   
  102   &   04-11    &        1188             &       27                      \\   
  103   &   04-12    &        1205             &       17                      \\   
  104   &   04-13    &        1212             &       7                       \\   
  105   &   04-14    &        1220             &       8                       \\   
  106   &   04-15    &        1248             &       28                      \\   
  107   &   04-16    &        1269             &       21                      \\   
  108   &   04-17    &        1305             &       36                      \\   
  109   &   04-18    &        1318             &       13                      \\   
  110   &   04-19    &        1331             &       13                      \\   
  111   &   04-20    &        1336             &       5                       \\   
  112   &   04-21    &        1345             &       9                       \\   
  113   &   04-22    &        1354             &       9                       \\   
  114   &   04-23    &        1367             &       13                      \\   
  115   &   04-24    &        1374             &       7                       \\   
  116   &   04-25    &        1389             &       15                      \\   
  117   &   04-26    &        1396             &       7                       \\   
  118   &   04-27    &        1402             &       6                       \\   
  119   &   04-28    &        1408             &       6                       \\   
  120   &   04-29    &        1418             &       10                      \\   
  121   &   04-30    &        1429             &       11                      \\   
  122   &   05-01    &        1434             &       5                       \\   
  123   &   05-02    &        1439             &       5                       \\   
  124   &   05-03    &        1439             &       0                       \\   
  125   &   05-04    &        1439             &       0                       \\   
  126   &   05-05    &        1445             &       6                       \\   
  127   &   05-06    &        1448             &       3                       \\   
  128   &   05-07    &        1449             &       1                       \\   
  129   &   05-08    &        1450             &       1                       \\   
  130   &   05-09    &        1454             &       4                       \\   
  131   &   05-10    &        1457             &       3                       \\   
  132   &   05-11    &        1460             &       3                       \\   
  133   &   05-12    &        1461             &       1                       \\   
  134   &   05-13    &        1463             &       2                       \\   
  135   &   05-14    &        1464             &       1                       \\   
  136   &   05-15    &        1465             &       1                       \\   
  137   &   05-16    &        1465             &       0                       \\   
  138   &   05-17    &        1466             &       1                       \\   
  139   &   05-18    &        1466             &       0                       \\
  140   &   05-19    &        1467             &       1                       \\
  141   &   05-20    &        1468             &       0                       \\
  142   &   05-21    &        1468             &       0                       \\
  143   &   05-22    &        1468             &       0                       \\
  144   &   05-23    &        1468             &       0                       \\
  145   &   05-24    &        1468             &       0                       \\
      \hline
  \end{tabular*}      
    \caption{Total Confirmed Cases and New Confirmed Cases per Day of COVID-19 Epidemic
      in Slovenia Reported by the National Institute of Public Health Data
      According to Wikipedia \cite{Covid19Slovenia_wikipedia}.}
      \label{table:slo}
  \end{center}
\end{table}

\end{document}